# A Blockchain-based Model for Securing Data Pipeline in a Heterogeneous Information System

Maneo Ntseliseng Ramahlosi[1], Yolo Madani[2] and Adeyinka Akanbi[3][0000-0002-8796-0674]

[1] Central University of Technology, Free State 9300, South Africa
MRamahlosi@cut.ac.za
[2] Central University of Technology, Free State 9300, South Africa
ymadani@cut.ac.za
[3] Central University of Technology, Free State 9300, South Africa
aakanbi@cut.ac.za

**Abstract.** In our digital world, access to personal and public data has become an item of concern, with challenging security and privacy aspects. Modern information systems are heterogeneous in nature and have an inherent security vulnerability, which is susceptible to data interception and data modification due to unsecured communication data pipelines between connected endpoints. This research article presents a blockchain-based model for securing data pipelines in a heterogeneous information system using an integrated multi-hazard early warning system (MHEWS) as a case study. The proposed model utilizes the inherent security features of blockchain technology to address the security and privacy concerns that arise in data pipelines. The model is designed to ensure data integrity, confidentiality, and authenticity in a decentralized manner. The model is evaluated in a hybrid environment using a prototype implementation and simulation experiments with outcomes that demonstrate advantages over traditional approaches for a tamper-proof and immutable data pipeline for data authenticity and integrity using a confidential ledger.

**Keywords:** Data security, Heterogeneous systems, Blockchain.

## 1 Introduction

Technological advancements have ushered in the Fourth Industrial Revolution (4IR) and the purported Fifth Industrial Revolution (5IR), with data becoming a critical asset in modern information systems. According to the statistics [7], [8], [9], 2.5 quintillion bytes of data are created each day, estimated to be worth around $77 billion by 2025 [1]. Data is the foundation of an information system, critical to the functioning, and plays a vital role in decision-making and strategic planning within an organization. Today, integrating data from heterogenous sources coupled with the integration of heterogeneous systems is an important capability for most organizations' successful decision-making, from data collection to processing and ultimately data analytics; It is critical that data accuracy and security be prioritized throughout the process. Observations based on empirical evidence indicate that the exchange of data across multiple application endpoints exposes the data to a high risk of manipulation and destruction by

malicious entities [2]. The attacks might have originated from any source, through a wide variety of attack mechanisms.

Typically, malicious hackers or state actors target accessible communication pathways of information systems – usually, data pipelines, which are not protected, using attacks such as man-in-the-middle (MITM) attacks, Cross Site Scripting (XSS), DNS Tunneling etc. The vulnerabilities of this critical communication channel have been the focus of researchers in information security [10]. Several countries have protection policies and fines for a data breach, urging the organization to provide rigorous mechanisms in place for data security. For example, South Africa promulgated the Protection of Personal Information Act, 2013 (Act 4 of 2013) ('POPIA') into law on 26 November 2013 and became fully enforceable on 1 July 2021 [11], with existing General Data Protection Regulation (GDPR) and other acts, resulting in, Equifax, a credit reporting agency agreeing to pay more than R10-billion to regulators to settle claims from a data breach [11]. Furthermore, if a company violates the General Data Protection Regulation (GDPR), the EU authorities may impose fines of up to 20 million euros or 4% of annual global revenue [5]. While these measures instigate the adoption of appropriate measures for adequate data security, they are not the solution.

Currently, traditional security mechanisms such as cryptographic techniques [3] are frequently insufficient to maintain data integrity in this massive modern IT infrastructure. Recently, researchers have turned their attention to blockchain, being a distributed, incorruptible, and tamper-resistant ledger database. Blockchain has the potential to address the critical security vulnerability issues of information systems, particularly on data integrity and reliability or in applications that require extra trust guarantees [4]. Blockchain technology can be used to secure data by creating a decentralized, tamper-proof ledger of transactions. Each block in the chain contains a cryptographic hash of the previous block, creating a chain of trust that is difficult to alter. By storing data on a blockchain, it becomes much more difficult for hackers to tamper with or steal the data, as any changes would be immediately apparent and could be easily traced back to the source. Additionally, the decentralized nature of blockchain technology means that no single point of failure exists, which further increases the overall security of the data (Fig. 1). This research explores how the application of blockchain technology could be used to mitigate the security challenges confronting heterogeneous information systems using environmental monitoring systems as a case study.

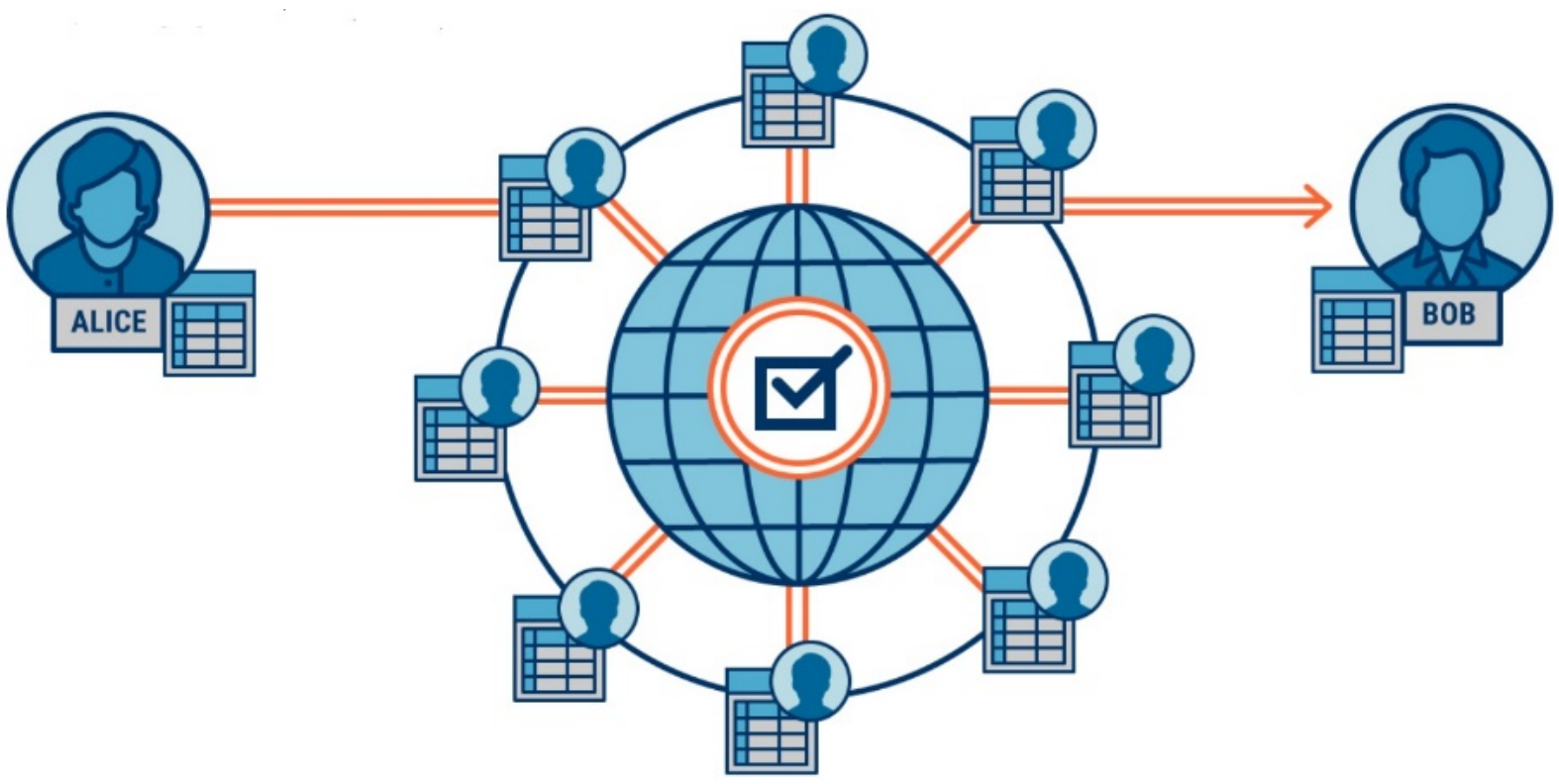

**Fig. 1.** A decentralized Blockchain ledger.

The environmental monitoring domain generates lots of critical and sensitive data that influences policies and strategy and is a prominent sector for the application of blockchain technology for data security. For example, an attacker could compromise the readings of the monitoring system, providing scientists with misleading data, if this information informs policy changes or new technologies, it could thoroughly disrupt environmental efforts. In such a scenario, the application of blockchain technology can be the more favourable approach to attain security from recognized threats and mitigate against possible data compromise. Blockchain can also be used to secure heterogeneous data from environmental monitoring systems. By storing data on the blockchain, it can be ensured that the data is accurate and cannot be tampered with. This can help to ensure that environmental monitoring systems are functioning properly. Blockchain technology secures data using advanced encryption methods, making it nearly impossible for unauthorized parties to access data stored on the blockchain on decentralized multiple nodes [12]. Decentralization is a core feature of blockchain technology, and it refers to the distribution of data and controls across a network of nodes, rather than having a central point of control – as a distributed ledger. Blockchain networks that are distributed among participants are called a consortium network. The consortium network gives each partner visibility into every transaction that occurs on the network. In a decentralized blockchain network, all nodes have a copy of the same data, containing an immutable ledger of all transactions. Each node has the ability to validate and record transactions making it easy to track and trace any changes made to the data.

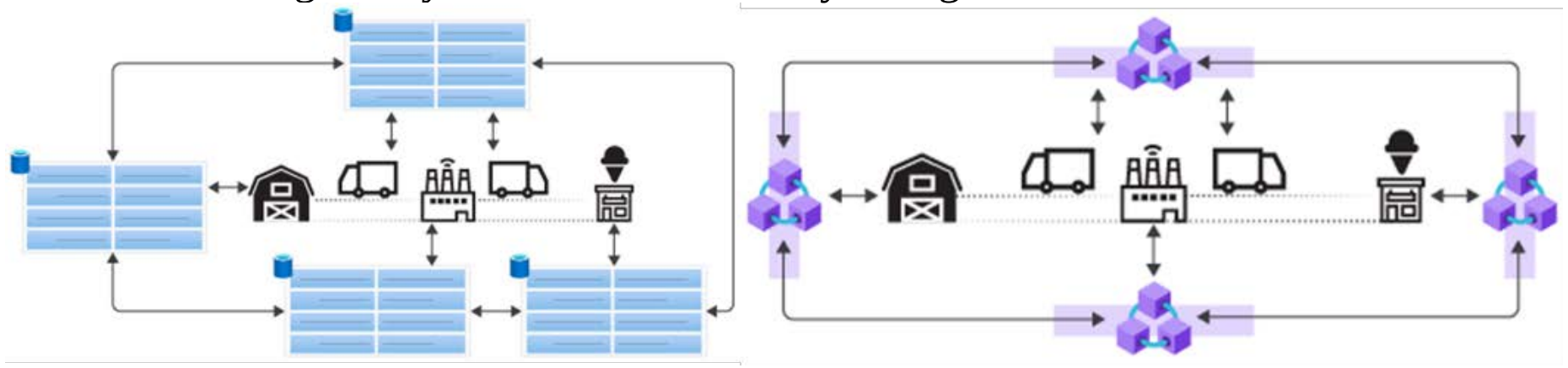

**Fig. 2.** Overview of a distributed database in comparison with a distributed ledger in blockchain

Blockchain makes use of smart contracts, which are self-executing contracts that can be programmed to automatically enforce data security protocols, such as terms of an agreement between data producer and data consumer written into lines of code. Smart contracts can be written and executed on blockchains like Ethereum [13]. These contracts are transparent and safe since they are created using code and kept on the blockchain. Smart contracts extend blockchain from data to code. One of the main advantages of smart contracts is their ability to automate and streamline procedures, eliminating the need for middlemen and resulting in time and cost savings. They also offer some amount of security and confidence due to the public availability and verifiability of the contract's code. In this area, numerous applications have been proposed, including hyper ledger fabric technology [14], Ethereum smart contracts [13], and other Blockchain as a Services (BaaS) [15, 16].

The goal of this research study is to explore how the application of how blockchains could be integrated with existing legacy and modern information systems. The main contributions of this paper are as follows:

- It proposes a decentralized blockchain architecture for securing data pipelines of heterogeneous information systems from attacks using a multi-hazard early warning system (MHEWS) as a case study.
- The proposed framework is implemented in the cloud using Azure Confidential Ledger – a ledger service that provides the ability to scale and operate blockchain networks on the Azure infrastructure.
- The effectiveness of our proposed framework is demonstrated and determined through the implementation and deployment of the Blockchain model in the cloud.

The rest of the paper is organized as follows, section 2 discusses related works, and section 3 outlines the application scenario with the presentation of the proposed model. Section 4 of this paper covers the methodology and approach for the experimental setup – for the implementation and deploying using the case study MHEWS, and section 5 provides the conclusion and future work.

## 2 Related Works

In the past few years, there has been a significant increase in the use of blockchain technology for security and privacy in information systems, the Internet of Things (IoT), healthcare, and cloud services. This growth is largely attributable to blockchain's capacity to secure data using cutting-edge encryption techniques, decentralization, Immutable Ledger, and Smart Contracts. Reference [20] developed a model for ICT e-agriculture systems with a blockchain infrastructure for use at the local and regional scale, with a focus on the specific technical and social requirements of blockchain technology for protecting ICT e-agriculture systems. References [21] investigate the intentional use of blockchain technology for data validation, data storage, data security, and data transfer to create decentralized, effective, fault-tolerant, and interoperable e-agriculture information systems. Reference [22] suggests a paradigm that makes use of smart contracts to guarantee reliable communication between devices and sensors. The authors of [23] conducted a survey that provides comprehensive reporting on various blockchain studies and applications put forth by the research community, as well as their respective effects on blockchain and its use across other applications or scenarios; their findings showed that blockchain is increasingly being used in contemporary cloud- and edge-computing paradigms.

The authors of [6] investigate how blockchain technology can be used for environmental monitoring, with a focus on key metrics like the quantity of articles published, author collaboration networks, research institution networks, and keyword co-occurrence in this area over an eight-year period. Some of the nine elements utilized to assess the literature included the number of papers published, author collaboration networks, research institution networks, keyword co-occurrence, co-citation analysis, keyword clustering, keyword burst, time zone, and the distribution of countries and regions. The study used Citeseer to visually explore the elements of cross-over literature studies, the characteristics of the collaborations, the influence of the contributions, etc. in addition to describing the qualities indicated by the data and came to a conclusion on the wide adoption of blockchain technology in different research fields (Fig. 3).

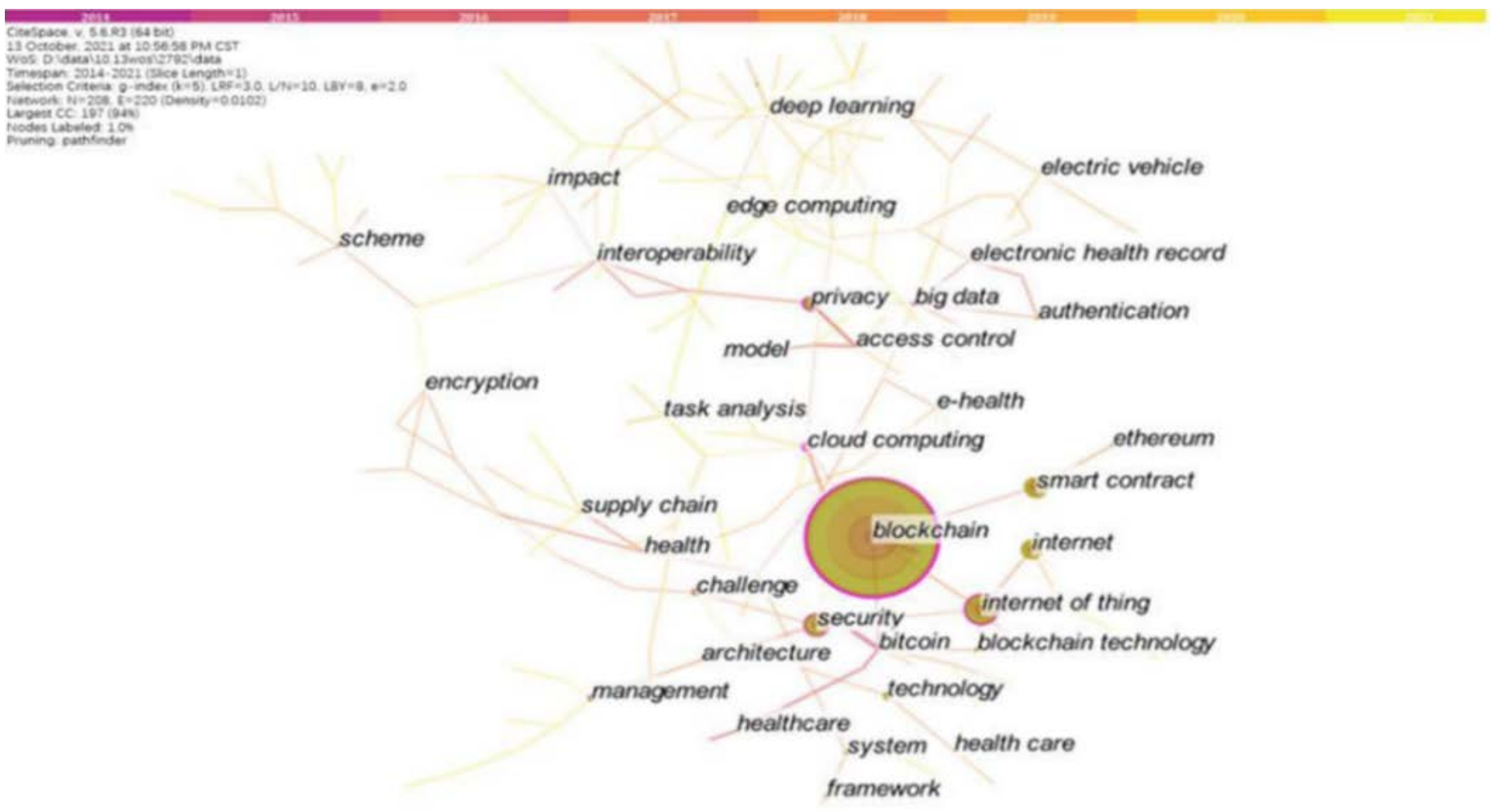

**Fig. 3.** Keyword concurrence [6].

## 3 Application Scenario and Proposed Model

In this section, we introduced the case study application scenario of an MHEWS in the environmental monitoring domain and the architecture of heterogeneous information systems from existing legacy systems, wireless sensor networks (WSN) etc., all integrated together for a comprehensive monitoring system.

### 3.1 Application Scenario

Information systems in the environmental monitoring context are heterogeneous by nature [17,18], with interoperability a major challenge [24]. Hence, components are isolated and integrated for processing based on the ETL model. There are three identified parts in the application scenario as shown depicted in Fig. 4 below, data producers, data pipeline and data consumers. Each part of the application scenario is integrated with other parts. The data producers in this case study of the integrated multi-hazard early warning systems (MHEWS) are the legacy systems, WSN and also standalone applications. The data pipeline consists of communication channels, which are either wired or wireless based on the infrastructure design. The data consumers are the endpoint application utilizing the data from the sensors and enterprise systems. The details of these three parts are illustrated below:

**Data Producers.** In a typical MHWES, in-situ readings produced by different producers (sensors and devices) are incorporated into accompanying EWS in various structured formats and types are streamed via the communication channel to the repository for further processing. In this case, examples of the parameters measured are temperature, pressure, moisture, humidity etc.

**Data Pipeline.** This layer represents the entire communication channel of the MHEWS. The communication channel could be wired or wireless utilizing several communication protocols like WiFi, Bluetooth, ZigBee etc. Irrespective of the communication medium adopted, data are susceptible to several attacks in this layer. The common encryption method offered by the communication protocols is not strong enough to prevent data compromise.

**Data Consumers.** The consumers are software programs or tools that are designed to process and analyze data in order to extract valuable insights or information. These applications can be used for a wide range of purposes, including business intelligence, machine learning (ML), predictive analytics, and data visualization.

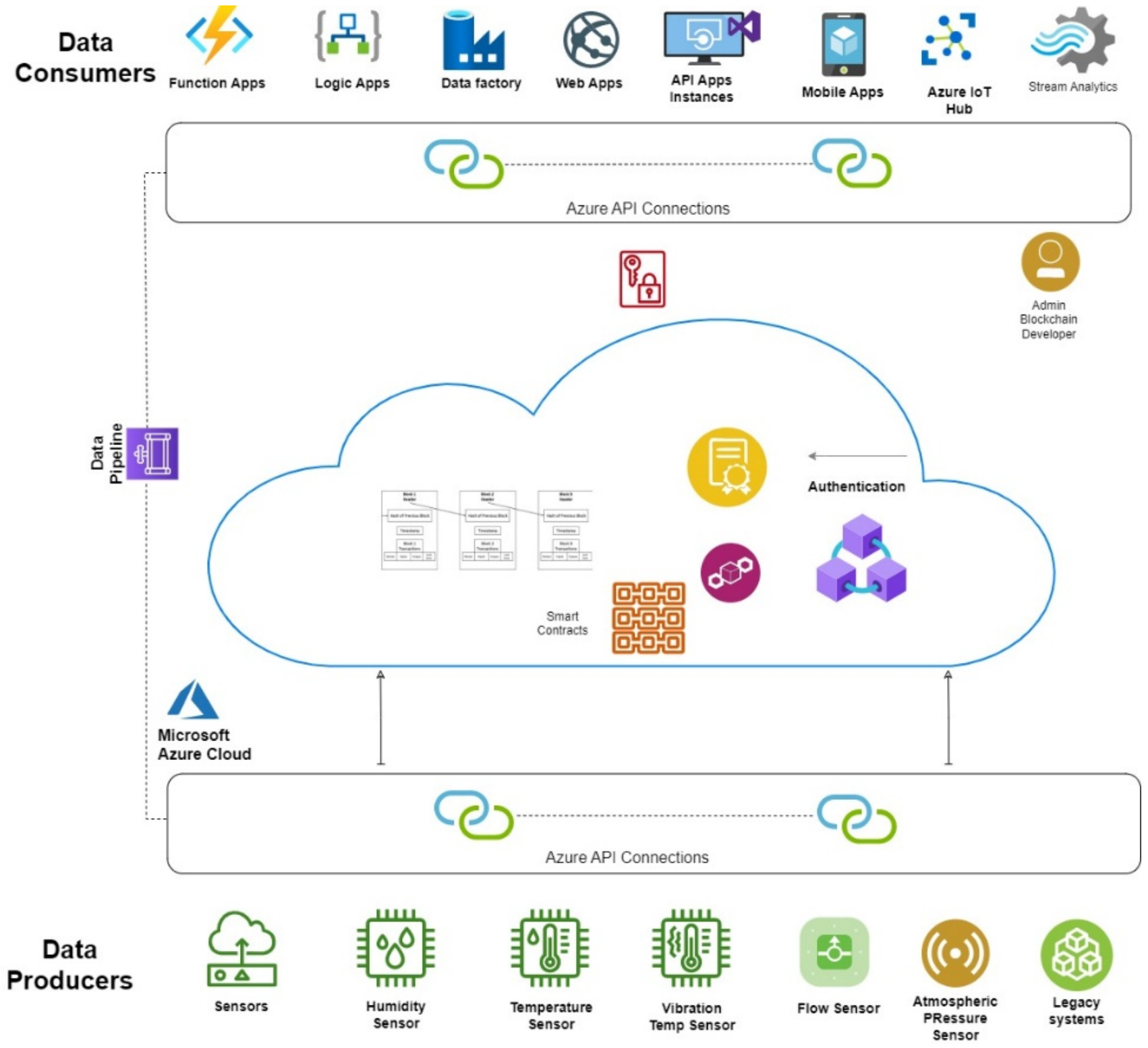

**Fig. 4.** An overview of an integrated Multi-hazard Early Warning System Architecture.

### 3.2 Proposed Blockchain-based Data Security Model

This paper proposes a Blockchain-based data security model (Fig 5) that applies blockchain technology in securing the data pipeline of an integrated multi-hazard early warning system, from the data producers to the data consumers. Each process and layer of our proposed model is described, outlining the data flow as data are generated, encrypted and validated in a hybrid model approach. The data obtained from the heterogenous data consumers in real-time are transmitted to the cloud using appropriate APIs and are added to the blockchain infrastructure in accordance with the smart contract. The data is cloned and replicated across all the nodes in the distributed database network of the blockchain [25]. Therefore, a compromise of the data accessed in one node does not affect the accuracy of the entire dataset in the cluster. The validation that the retrieved data is immutable is achieved by cross-referencing with the blocks stored on the ledger and all user transactions.

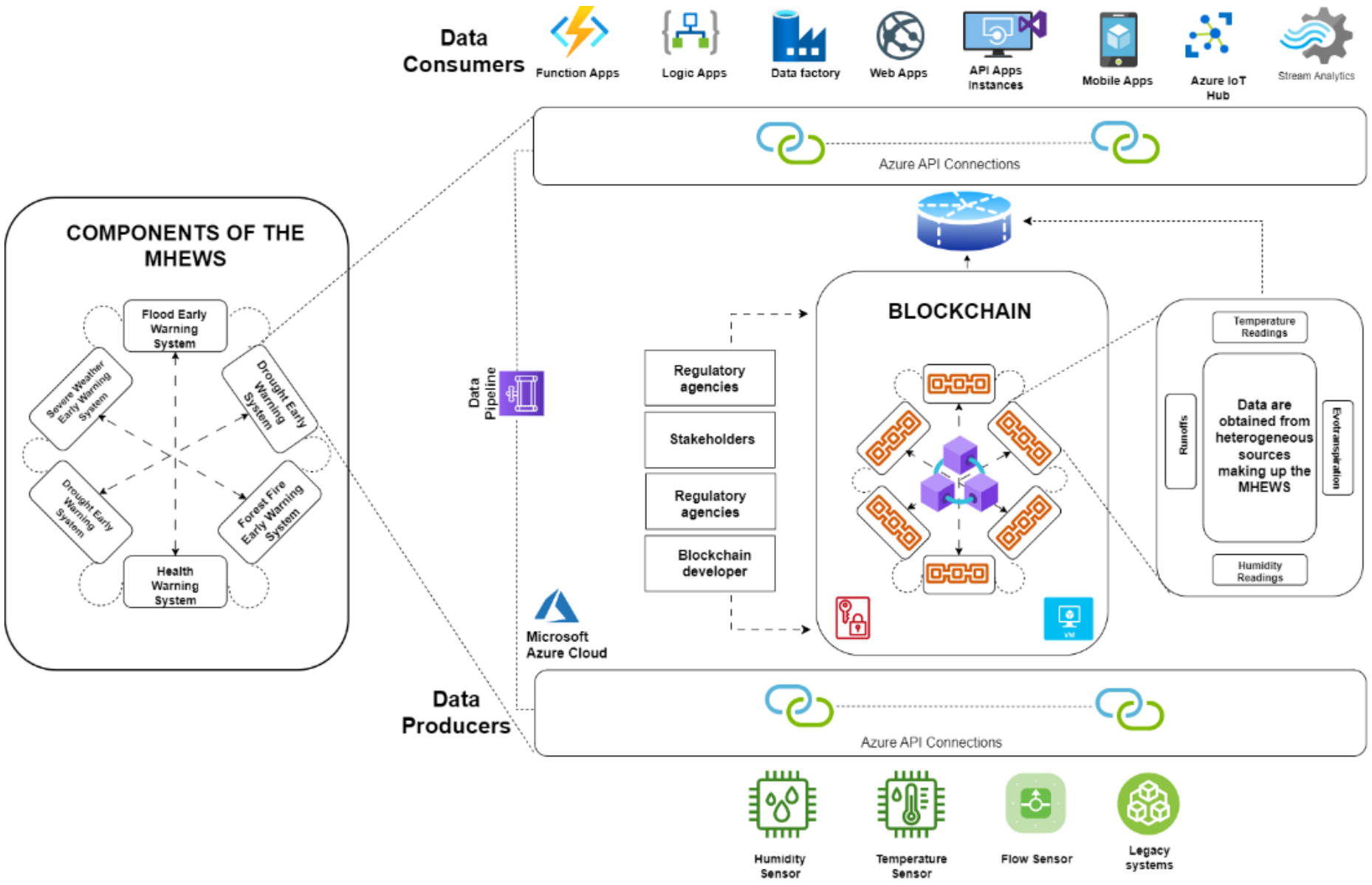

**Fig. 5.** A Blockchain-based Data Security Model for an integrated Multi-hazard Early Warning System (MHEWS)

## 4 Methodology

The implementation of blockchain technology for data security can be a complex process, herein the process adopted are simplified and broken down into several steps:

I. Define the use case: This first step involves the identification of the specific data or information that needs to be secured and how it will be used, followed by the specific business logic and conditions that need to be encoded into the smart contract. In the study - case of environmental monitoring systems, the minimum and maximum readings of the measured parameters are noted as conditions/rules encoded into the smart contract. For example, early warning indicators such as temperature can have low and high values of -50celsius to +50celcius respectively.

II. Choose a blockchain platform: There are several blockchain platforms that supports smart contract functionality, such as Ethereum or EOS with functionality that offers a complete end-to-end solution for developing, hosting, and managing blockchain solutions. In a hybrid model design that allows data transmission to an online repository, the use of such services in Azure Cloud Security, which is best in class keeps both processes and data secure [19]. Azure Blockchain service supports Ethereum, Quorum Ledger, Corda, and Hyperledger Fabric.

This research employs Microsoft Confidential Ledger (ACL), a managed and decentralized ledger for data entry that is supported by Blockchain. Data committed to the Confidential Ledger is made tamperproof in perpetuity by ACL through consensus-based replicas and cryptographically signed blocks, prohibiting intentional or unintentional data alteration.

III. Write the smart contract code: Blockchains, such as Ethereum, allow for the creation and execution of smart contracts. These contracts are written in code and stored on the blockchain, making them transparent and secure. The business logic and conditions are coded using a programming language, such as Solidity.

IV. Test and deploy the smart contract: Test the smart contract using a virtual environment, such as a Remix IDE, to ensure that it functions as intended. Deploy the smart contract to the blockchain network, making it available for execution.

V. Monitor and maintain: Monitor the network for any security breaches and perform regular maintenance to ensure the network remains secure.

In this case study, the ACL is created in the cloud, based on the proposed hybrid architecture. In the cloud platform, the model is implemented, starting with the creation of the resource group that will contain the resources and services required for the implementation and deployment of the infrastructure. The ability to implement, deploy and scale are considered one of the main advantages of cloud services. Using a component of the MHEWS, drought early warning systems (DEWS) services are replicated in the cloud platform, starting with the configuration of the in-situ sensors and legacy systems to transmit environmental readings to the cloud repository using ledger APIs support certificate-based authentication process with owner roles as well as Azure Active Directory (AAD) based authentication and also role-based access. The data pipelines from the endpoints to the ledger are transmitted over a TLS 1.3 connection, which terminates within the hardware-backed security enclaves (Intel® SGX enclaves).

The secure ledger is accessible via REST APIs, which can be integrated into new or existing applications using JSON resource connection strings as captured in a code snippet (Fig. 6) and Azure service in Fig. 7 below. The data streams are created as blocks in blob storage containers belonging to an Azure Storage account in an encrypted format and committed to the ledger. Basic operations such as create, update, get, and delete are supported by the Administrative APIs.

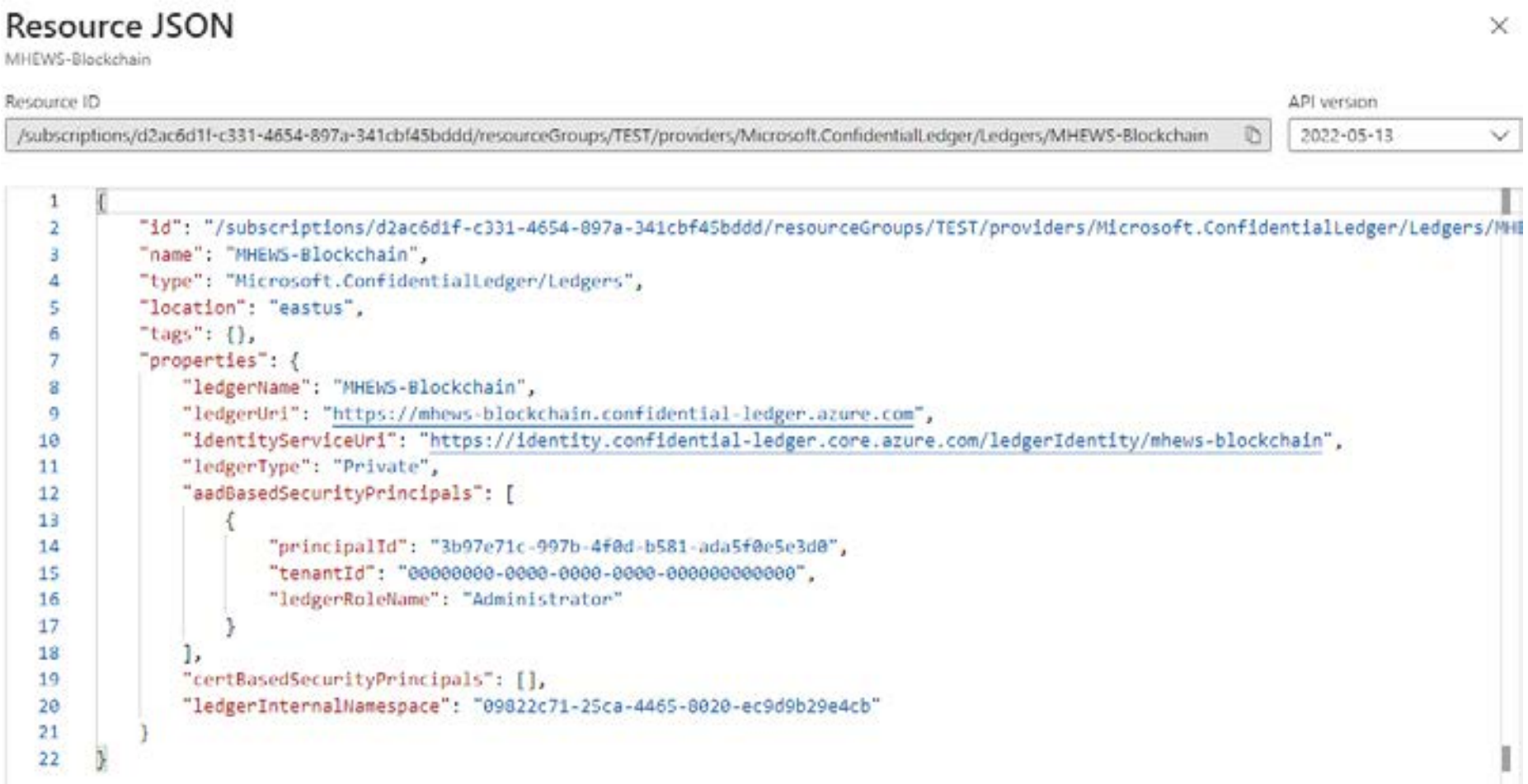

**Fig. 6.** Code Snippet – A JSON REST API for accessing the MHEWS ACL in the Azure.

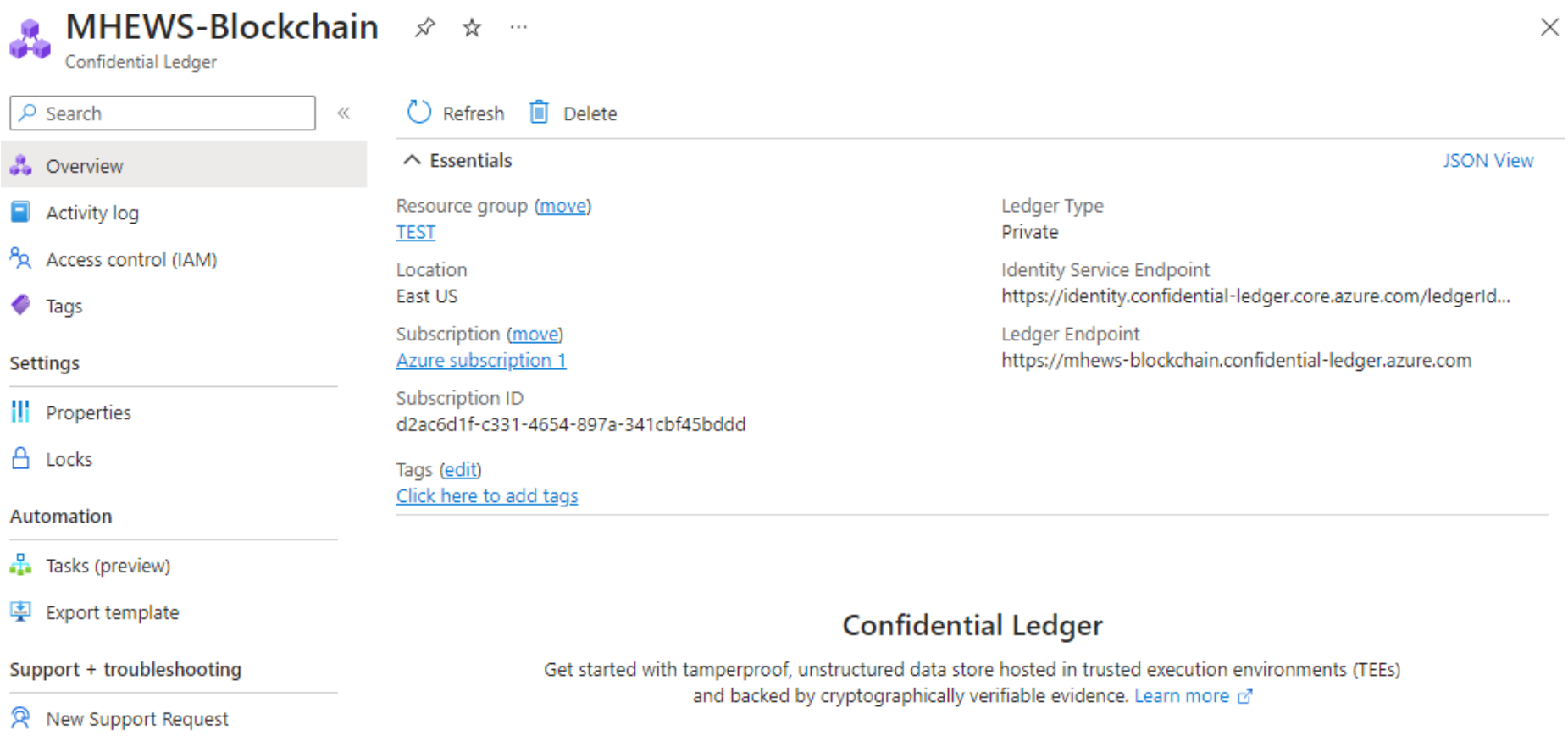

**Fig. 7.** The MHEWS Blockchain service in Azure Cloud.

Smart contract manages the nodes to govern the consensus in the ledger. Smart contracts are implemented on the blockchain to automate the process of data validation, reducing the potential for human error or manipulation. Typically, the code of a smart contract is composed in a high-level programming language such as Solidity before being compiled into bytecode and uploaded to the blockchain. The smart contract that can be deployed on our blockchain will be written in Solidity and Remix. However, for test purposes before deployment, Truffle provides a testing framework and asset pipeline that makes it easy and quick to deploy contracts, develop applications, and run tests on a local server using tools like Ganache. Using the Temperature parameter from the deployed sensors of the DEWS, a smart contract is created for experimental purposes and deployed on the local server using Remix as captured in Fig 8 below.

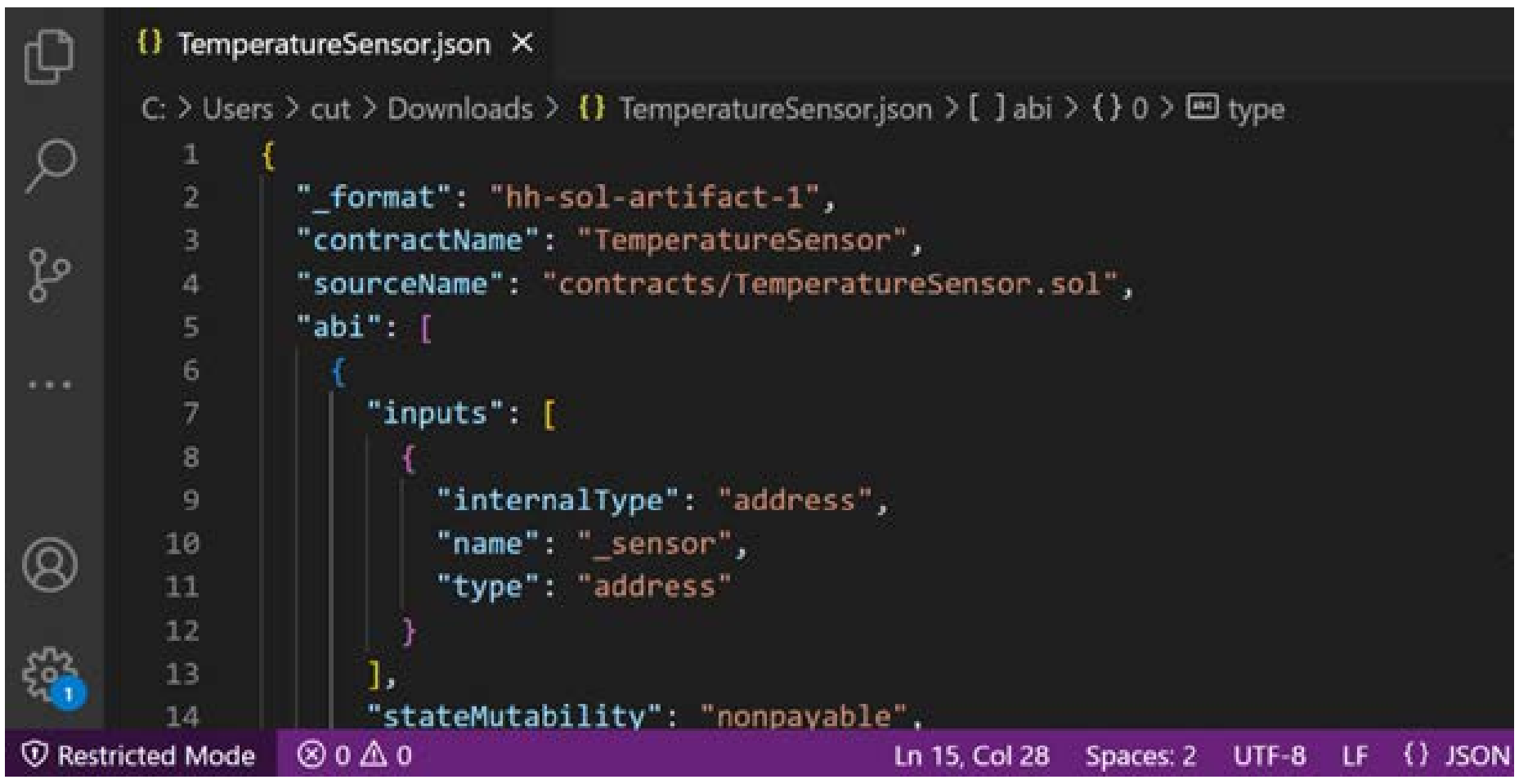

**Fig. 8.** Using the *Temperaturesensor.sol* smart contract as an example in Remix for testing purposes.

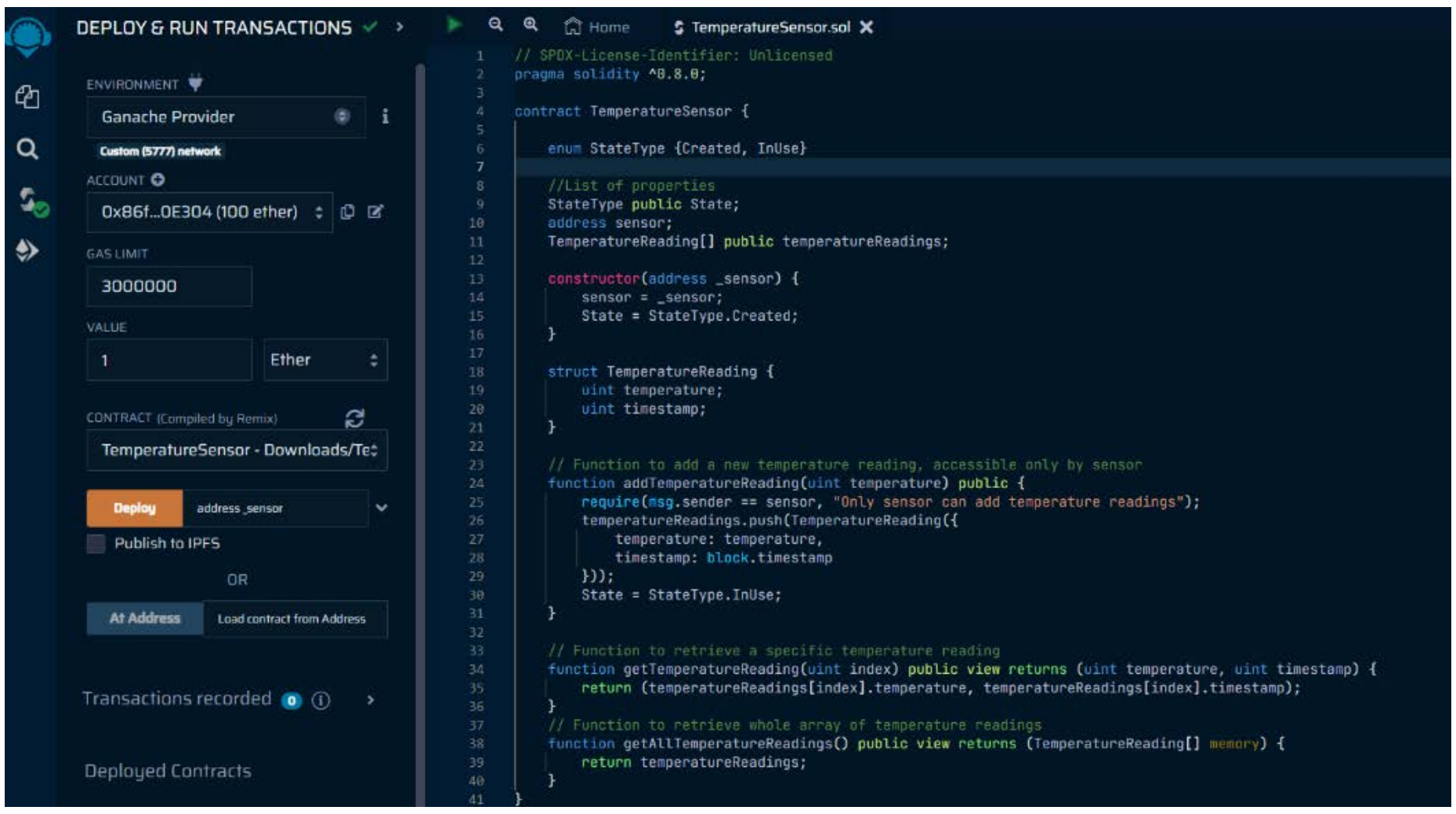

**Fig. 8.** The smart contract is deployed and tested on a local server before deployment to the cloud in Remix testing environment.

Publishing and using the smart contract created for the parameters of the reading involves copying the ledger endpoints to the Remix and selecting a Web3 Provider using the generated Azure ledger URL. Once connected, the contract is deployed, and interaction with the contract is achieved through Remix for testing purposes.

## Conclusion

Blockchain is a decentralized, distributed ledger that eliminates the need for a trusted third party or single point of failure. The presented blockchain security model in the experimental setup demonstrated has guaranteed the security of data pipelines in heterogeneous information systems such as the MHEWS case study. This model has the ability to work with new or existing applications similar to any PaaS, and IaaS models, using cloud-based apps and storage. The ACL in our Blockchain system creates an ever-growing list of ordered information called blocks to create a digital ledger. This list is made possible by a decentralised, peer-to-peer network. Each transaction is then automatically verified by the network itself after being included in a block with a cryptographically signed signature. This enables the security of environmental monitoring data by creating a decentralized and tamper-proof system for storing and sharing data.

The use of blockchain can ensure that the data is accurate, and reliable and cannot be altered without leaving a traceable record. Overall, the presented architecture offers distinct benefits for data integrity, such as immutability, tamper-proofing, and append-only operations in the relevant domain. These capabilities, which guarantee the preservation of all records, are perfect when it's necessary to preserve essential metadata records, such as for archive and regulatory compliance needs. As outlined for future work will focus on the performance metrics of implementing the model in a hybrid model and possibility of containerization or deployment of the MHEWS components as a microservice.